\documentclass[prb,twocolumn,showpacs,preprintnumbers,amsmath,amssymb]{revtex4}
\usepackage{graphicx}
\usepackage{dcolumn}
\usepackage{bm}
\usepackage{psfig}
\newcommand \be{\begin{eqnarray}}
\newcommand \ee{\end{eqnarray}}

\begin{document}
\title{Phase diagram for interacting Bose gases}
\author{K. Morawetz$^{1,2}$, M. M{\"a}nnel$^{1}$ and M. Schreiber$ ^1$}
\affiliation{$^1$Institute of Physics, Chemnitz University of Technology, 
09107 Chemnitz, Germany}
\affiliation{$^2$Max-Planck-Institute for the Physics of Complex
Systems, N{\"o}thnitzer Str. 38, 01187 Dresden, Germany}

\begin{abstract}
We propose a new form of the inversion method in terms of a selfenergy expansion to access the phase diagram of the Bose-Einstein transition. The dependence of the critical temperature on the interaction parameter is calculated. This is discussed with the help of a new condition for Bose-Einstein condensation in interacting systems which follows from the pole of the T-matrix in the same way as from the divergence of the medium-dependent scattering length. A many-body approximation consisting of screened ladder diagrams is proposed which describes the Monte Carlo data more appropriately. The specific results are that a non-selfconsistent T-matrix leads to a linear coefficient in leading order of $4.7$, the screened ladder approximation to $2.3$, and the selfconsistent T-matrix due to the effective mass to a coefficient of $1.3$ close to the Monte Carlo data. 
\end{abstract}
\date{\today}
\maketitle

\section{Introduction}
Interacting Bose gases have become a topic of current interest triggered by the experimental table-top demonstration of Bose condensations  [\onlinecite{AWMEC95,DMADDKK95}]. Especially, it is of great importance to know the behavior of the critical temperature in dependence on the interaction strength and the density. The fluctuations beyond the mean field are especially significant since they establish the deviations of the critical temperature from the noninteracting value and might be observable by adiabatically converting down the trapping frequency  [\onlinecite{HSC97}]. The measurement of the critical temperature of a trapped weakly interacting $^{87}$Rb gas differs from the ideal gas value already by two standard deviations [\onlinecite{GTRHBA04}]. Therefore it is of great interest to understand this change even in bulk Bose systems. 

The density $n_0$ of the noninteracting bosons with mass $m$ and temperature $T$ reads
\be
n_0(T)=\left({m k_B T \over 3 \pi \hbar^2} g(\epsilon)\right )^{3/2}
\label{n0}
\ee
where
$g(x)=\frac 3 2 [P_{3/2}(x)]^{2/3}$ is given in terms of the polylog function $P_{3/2}(x)={2\over \sqrt{\pi}} \int\limits_0^\infty {\sqrt{t} dt\over {\rm e}^{t+x}-1}$ and we denote the negative ratio of the chemical potential to the temperature  by $\epsilon=-\mu/k_B T$. The critical temperature $T_0(n)$ is given by (\ref{n0}) for $\epsilon\to 0$.

The critical temperature, $T_c$, of the Bose system with repulsive interaction is deviating from the free one, $T_0$, and is a function of the free scattering length $a_0$ times the third root of the density. One usually discusses this deviation on condition that the density of the interacting system at the critical temperature, $n(a_0,T_c)$, should be equal to the density of the noninteracting Bose system $n_0(T_0)$ at the critical temperature $T_0$.
According to (\ref{n0}) we have $n_0(T_c)=(T_c/T_0)^{3/2} n_0(T_0)$ and one obtains  [\onlinecite{BBHLV99}]
\be
x={T_c\over T_0}=\left (n_0(T_c)\over n(a_0,T_c)\right )^{2/3}
. 
\label{x0}
\ee

The change of the critical temperature ${(T_c-T_0) / T_0}$ in dependence on the coupling strength $y_0=a_0 n^{1/3}$ has been found to be linear in leading order. Monte-Carlo (MC) simulations and theoretical calculations by many groups have confirmed
\be
{T_c-T_0 \over T_0}\approx c_1 \, y_0 +{\cal O}(y_0^2).
\ee
The sometimes occurring square root dependence has turned out to be an artifact of the first order virial expansion [\onlinecite{BBHLV01}]. Although there is now an agreement about the linear slope, the actual value of $c_1$ varies from 0.34 to 3.8 as discussed in [\onlinecite{A04,KSP04,BBHLV01}]. The situation is different for finite Bose systems contained in a trap. The increase of interaction lowers the density in the center of the trap and reduces effectively the critical temperature  [\onlinecite{GTRHBA04}] such that $c_1=-0.93$ was found in [\onlinecite{WIK00}]. Here we want to consider only bulk Bose gases.

Let us shortly sketch the different expansion schemes. The convergence of such expansions is discussed in [\onlinecite{BR02}] in terms of the large-$N$ expansion of scalar field theory. The optimized linear expansion method  [\onlinecite{CPR01}] yielded $c_1=3.06$ for two-loop contributions and $c_1=1.3$ for six loops [\onlinecite{KNP04}]. Though this $\delta$-expansion works well in quantum mechanics it fails to converge in quantum field theory [\onlinecite{HK03}]. The
$1/N$ expansion  [\onlinecite{BBZ00}] provides $c_1=2.33$. This seems to be in agreement with the non-selfconsistent summation of  bubble diagrams [\onlinecite{BBHLV01}] which gives $c_1={8\pi \over 3}\zeta(3/2)^{-4/3}\approx2.33$. The same result has also been found with a T-matrix approximation  [\onlinecite{PST01}] . Older MC data [\onlinecite{HK99}] show similar values such as $c_1=2.30\pm0.25$ while newer ones, [\onlinecite{AM01,KPS01}], report $c_1\approx1.3$. The latter result can also be found by variational perturbation theory [\onlinecite{K03}] yielding $c_1=1.23\pm0.12$. The same coefficient has been obtained by exact renormalization group calculations [\onlinecite{LHK04,HLK04}] which provide the momentum dependence of the self energy at zero frequency as well. The usage of Ursell operators  [\onlinecite{HGL99}] has given an even smaller value $c_1=0.7$.

Though the weak coupling behavior of the critical temperature can be considered as settled there is considerably less known about the strong coupling behavior. In [\onlinecite{KSP04a}] a nonlinear behavior of the critical temperature was proposed; first the interaction leads to an increase of the critical temperature up to a maximal one. A further increase of interaction strength or correlations then decreases the critical temperature until the Bose condensation vanishes at a maximal coupling parameter $y_0^{\rm crit}$. One finds the interesting physics that too strong repulsion prevents the Bose condensation.

The variational perturbation method developed in  [\onlinecite{K04}] and advocated in [\onlinecite{KSP04a}] allows one to sum up higher order diagrams in order to account for strong coupling effects. In this paper we present an alternative method well known as inversion method to obtain from the ladder summation diagrams a higher order perturbation result by inversion. The motivation to propose yet another theoretical method besides the already mentioned Monte-Carlo simulations and the variational perturbation method is twofold. First the strong coupling limit is a problem of theoretical challenge which is not yet solved. Different schemes favor different summations of diagrams and therefore weight different processes differently. We think that different paths and comparing the results between them might be enlightening for understanding the physical mechanisms. The second motivation is a pure methodological one. The proposed inversion method has led already to remarkable results in other fields. Therefore we think it is interesting to try how far this method can be applied fruitfully to the highly nontrivial problem of interacting Bose gases. We expect from this method theoretical insights into the systematics of higher order processes which have to be considered for the description of strongly coupled systems showing spontaneous symmetry breaking. Here the inversion method seems to have its advantages since it allows explicitly for symmetry breaking. This is, however, beyond the scope of the present work.   

Below we discuss first the method to extract the phase diagram within the standard ladder approximation solving the Feynman-Galitzky form of the Bethe-Salpeter equation. This will be performed with the help of the separable potential which is then used in the limit of the contact interaction. This way is mathematically convenient since we can renormalize the coupling strength to the physical scattering length avoiding divergences. Then we apply the inversion method starting from the ladder approximation. This leads us to a twofold limit of the resulting series, one of which shows the nontrivial behavior in the phase diagram due to symmetry breaking. We compare this result with the Monte Carlo data and other theoretical approaches in the literature. 
In the last chapter we discuss improvements, mainly the summation of ring diagrams connected with ladder diagrams which will modify the linear slope but will not change the scaled nonlinear curve. Possible further improvements are outlined in the summary.

\section{Inversion method}

The inversion method we use has been discussed thoroughly in the centennial review paper of Fukuda {\em et al.}  [\onlinecite{FKYSOI95}]. We consider a system which can undergo a symmetry breaking due to a small external perturbation such that a phase transition is possible. The corresponding order parameter $\phi$ shall depend on an external source parameter $J$. We can expand the order parameter in terms of the coupling $g$ to the source via
\be
\phi=f(J)=\sum\limits_{n=0}^\infty g^n f_n(J)
\label{phi}
\ee
which corresponds to a perturbation series. Since we can perform only a calculation to finite order in $g$, a vanishing $J$ means also a vanishing order parameter $\phi$ and we are not able to describe the phase transition as a proper spontaneously broken symmetry. The situation can be improved if (\ref{phi}) 
is inverted 
\be
J=h(\phi)=\sum\limits_{n=0}^\infty g^n h_n(\phi).
\label{J}
\ee
For any expansion with a finite number of $f_n$ we can calculate the inverted coefficients $h_n$ by introducing (\ref{J}) into (\ref{phi}) and comparing the powers of $g$ to obtain  [\onlinecite{FKYSOI95}]
\be
h_0(\phi)&=&f_0^{-1}(\phi)
\nonumber\\
h_1(\phi)&=& - \left . {f_1(J)\over f_0'(J)}\right |_{J=h_0(\phi)} \qquad {\rm etc}.
\ee
If we now set $J=0$ in the inverted series, the nontrivial solution $\phi\ne 0$
is obtainable though with finite-order calculation of the $h_n$ in (\ref{J}).
The finite truncation in the inverted series (\ref{J}) corresponds to an infinite series summation in (\ref{phi}). For further details and successful applications of this method see [\onlinecite{FKYSOI95}].

We will use the ladder summation to establish a sum for the scattering length and will invert this series with respect to the selfenergy in order to obtain a nonlinear phase transition curve, i.e. the dependence of the critical temperature on the coupling strength.

\section{Many-body T-matrix approximation and Bethe-Salpeter equation}

\begin{figure}
\psfig{file=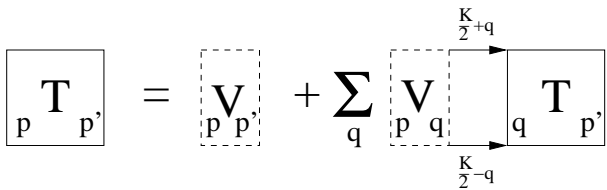,width=8cm}
\caption{The diagrams of the T-matrix approximation where the difference momenta are indicated as well as the momenta of the two intermediate propagators.
\label{tmat1}}
\end{figure}

The T-matrix depends on the frequency $\omega/\hbar$ and the center-of-mass momentum $K$ and reads according to figure \ref{tmat1} 
\be
T_{pp'}(\omega,K)\!=\!V_{pp'}\!+\!\sum\limits_{q} V_{pq} {1\!+\!f_{\frac K 2 \!-\!q}\!+\!f_{\frac K 2 \!+\!q}\over \omega\!-\!\varepsilon_{\frac K 2 \!-\!q}\!-\!\varepsilon_{\frac K 2 \!+\!q}\!+\!i \eta} T_{qp'}(\omega,K)
\nonumber\\
\!\!
\label{t-mat}
\ee
where we indicate the difference momenta of the incoming and the outgoing channel by subscript, the energy dispersion is assumed to be $\varepsilon_p=p^2/2m$ and the Bose functions are denoted by $f_k$.

This Bethe-Salpeter equation can be solved by a separable potential $V_{pp'}=\lambda g_pg_{p'}$ with a coupling constant $\lambda$ and a Yamaguchi form factor $g_p=1/(p^2+\beta^2)$ where $\beta$ describes the finite range of the potential. One obtains
\be
T_{pp'}(\omega, K)&=&{\lambda g_p g_{p'}\over 1-\lambda J(\omega, K)}
\nonumber\\
J(\omega,K)&=&\sum\limits_{q} g_{q}^2 {1+f_{\frac K 2 -{q}}+f_{\frac K 2 +{q}}\over \omega-\varepsilon_{\frac K 2 -{q}}-\varepsilon_{\frac K 2 +{q}}+i \eta}.
\label{JJ}
\ee
The scattering phase 
\be
\tan{\delta}=\left . {{\rm Im} T\over {\rm Re} T}\right |_{\scriptsize \begin{array}{l}p=p',K=0
\cr \omega ={K^2\over 4m}+{p^2\over m}
\end{array}} 
\approx a k+{\cal O}(k^2)
\label{adef}
\ee
is linked to the scattering length by the small momentum expansion, $p=\hbar k$. From (\ref{JJ}) we obtain   [\onlinecite{SMR97}]
\be
a=-{{\pi \hbar \lambda\over 2 \beta^4} (1+2 f_0)\over {2 \pi^2\hbar^3\over m}+\lambda \int\limits_0^{\infty} d{\bar p} {1+2 f_{\bar p}\over (\beta^2+{\bar p}^2)^2}}.
\label{a}
\ee
Now we can perform the limit to the contact potential using a $\beta$-dependent coupling constant $\lambda(\beta)$ in such a way that the free two-particle scattering length $a_0$ in the absence of many-body effects, $f_k\to0$, is reproduced  [\onlinecite{SMR97}]. From (\ref{a}) we have
\be
\lambda(a_0,\beta)=-{{2\pi^2 \hbar^3 \over m}\over {\pi \hbar\over 2 a_0 \beta^4}+\int\limits_0^{\infty} d{\bar p} {1\over (\beta^2+{\bar p}^2)^2}}
\label{lb}
\ee
which we use to eliminate $\lambda$ in (\ref{a}) to obtain $a(a_0,\beta)$. Performing now the limit $\beta\to\infty$ we obtain a well defined expression for the scattering length with many body effects
\be
{a\over a_0}={1+2 f_0\over 1-{4 a_0 \over \pi \hbar}\int\limits_0^\infty d{\bar p} f_{\bar p}}.
\label{aa0}
\ee
This scattering length depends on the chemical potential and on the temperature, $a(\mu,T)$ via the Bose function $f_p$. This renormalization procedure is completely equivalent to a momentum cut-off used in [\onlinecite{PS00}]. Higher order partial wave contact interactions have been discussed in [\onlinecite{RF01}] which are important for constructing pseudopotentials  [\onlinecite{SSBD05}]. For further relations of the T-matrices in different dimensions see [\onlinecite{M02}].

In [\onlinecite{SMR97}] the expression (\ref{aa0}) was discussed and it was found that this medium-dependent scattering length diverges at the Bose-Einstein transition characterized by the critical density and temperature.
We will employ (\ref{aa0}) as a good measure of approaching the Bose-Einstein condensate. With the help of (\ref{n0}) we can rewrite (\ref{aa0}) into an equation for the dimensionless interaction parameter $y=a n^{1/3}$ versus the free
one $y_0=a_0 n_0^{1/3}$, 
\be
{y\over y_0}&=&{{\rm coth}{\left (\epsilon_\Sigma\over 2\right )}\over 1+4 y_0 g'(\epsilon_\Sigma)}
\label{sys}
\ee
on the condition $n(a_0,T_c)=n_0(T_0)$.
The ratio of the temperatures $x=T_c/T_0$ for interacting to noninteracting Bose systems follows from (\ref{n0}) and (\ref{x0}) as 
\be
x&=& {g(0)\over g(\epsilon_\Sigma)}
\label{sys1}
\ee
where $\epsilon_\Sigma=\epsilon+\Sigma(\epsilon_\Sigma,y_0)/k_BT$ denotes the functional dependence of the quasiparticle energy in the argument of the distribution function on the selfenergy.

\section{Bose-Einstein condensation border line}

\subsection{T-matrix approximation}
Our aim is to extract the phase diagram of $a_0$ in terms of $x=T_c/T_0$. A simple elimination of $\epsilon$ in the system of equations (\ref{sys}) and (\ref{sys1})  would not lead to an answer since the ladder approximation is a weak coupling theory  [\onlinecite{StB93}] and cannot lead to the proper description of strong coupling. The simple elimination $\epsilon=\epsilon(y)\to x=x[\epsilon(y)]$ would correspond to the direct series (\ref{phi}). We will present how the strong coupling limit can be reached in the next chapter. Here we derive a proper condition for the Bose-Einstein condensation border line. This will be performed in two ways, first by the observation of a diverging medium dependent scattering length and second by the condition of the appearance of a Bose pole in the T-matrix. Both conditions will lead to the same criterion.

Starting we take into account the condition of being as near as possible to the phase separation line. We use the fact that the medium-dependent scattering length diverges at this phase separation line such that we obtain for critical $\epsilon_\Sigma^{\rm crit}\to\tilde \epsilon$ from (\ref{sys}) and (\ref{sys1})
\be
{y_0}&=&{y\over {\rm coth}{\epsilon_\Sigma\over 2}-4 y g'(\epsilon_\Sigma)}\to -{1\over 4 g'(\tilde \epsilon)}\equiv y_0^c
\nonumber\\
x&=& {g(0)\over g(\tilde \epsilon)}
\label{sys2}
\ee
as the system of equations which gives the separation line in the phase diagram. The meaning of $\tilde \epsilon=-\mu(a_0,T_c)/k_BT_c+\Sigma/k_BT_c$ is the critical chemical potential and the selfenergy for the interacting Bose gas divided by the critical temperature.

The same condition (\ref{sys2}) for the formation of a condensate can be found by calculating the binding pole of the T-matrix (\ref{JJ}).
With the renormalization (\ref{lb}) the bosonic T-matrix (\ref{JJ}) for contact interaction reads
\be
T_{p p'}(\omega,Q) = - {4 \pi \hbar^2 a_0 \over m}{1 \over 1 + J(\omega,Q)}.
\label{JJci}
\ee
In the center of mass frame one finds
\be
J(\omega,0) &=& - {4 a_0 \over \pi \hbar} \int\limits_0^\infty dq {q^2  f_q \over q^2 - m \omega}
\nonumber\\&+& {a_0 \over \hbar}\left\{\begin{array}{c} \sqrt{-\omega m}\cr - i \sqrt{\omega m}\left(1 + 2 f_\omega \right)\end{array} \mbox{for} \begin{array}{c} \omega \le 0 \cr \omega>0\end{array}\right..
\label{bcond}
\ee
In the dilute limit, $f_q \to 0$, the free two-particle phase shift of (\ref{adef}) starts from zero for positive $a_0$ and from $\pi$ for negative scattering length indicating a bound state according to the Levinson theorem [\onlinecite{GW64}]. For attractive interactions, $a_0<0$, we have therefore a bound state pole at  $J(-E_b,0)=-1$ with the binding energy $E_b=\hbar^2/m a_0^2$. With the sign of the scattering length in (\ref{adef}) we follow the convention of [\onlinecite{GW64}] which
  has an opposite sign compared to [\onlinecite{F94}]. Generally odd numbers of bound states correspond to negative scattering lengths and even numbers to positive scattering lengths.

We are interested here in the Bose condensation, that means in a pole for repulsive interaction, $a_0>0$. In this case there is no bound state, but a Bose-condensation pole possible at $J(-E_b,0)=-1$. We can estimate
\be
J(\omega\!<\!0,0) \,> \,J(0,0) =  - {4 a_0 \over \pi \hbar} \int\limits_0^\infty dq f_q = 4 y_0 g'(\epsilon).
\label{ab} 
\ee
The T-matrix (\ref{JJci}) has therefore a Bose-condensation pole if
$
J(0,0) \le -1
$
which means in view of (\ref{ab}) that
\be
y_0 \ge - {1 \over 4 g'(\epsilon)}.
\ee
This is the same condition as (\ref{sys2}) and one can see that the existence of a pole in the T-matrix is equivalent to the existence of a Bose condensation. In other words this pole represents the medium dependence of the Bose condensation in T-matrix approximation.

\begin{figure}
\psfig{file=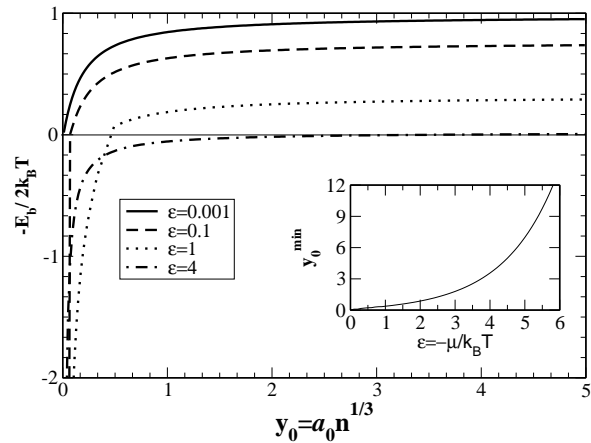,width=9cm}
\caption{The condition for appearance of a Bose-condensation pole, $J(-E_b,0)=-1$ according to (\protect\ref{bcond}). The Bose-condensation energy is plotted versus the dimensionless interaction parameter for different ratios of the chemical potential to the temperature, $\epsilon=-\mu/k_BT$. The inset shows the minimal interaction parameter as a function of $\epsilon$.\label{i_ey0}}
\end{figure}

Let us shortly discuss the physical content of the pole condition of the T-matrix.  The result according to (\protect\ref{bcond}) is plotted in figure~\ref{i_ey0} versus the dimensionless interaction parameter. For $\epsilon=-\mu/k_BT\to0$ the curve starts at zero which means that any small interaction leads to a positive Bose condensation energy while for noninteracting systems the condensation energy remains at zero. For the critical $\epsilon=0$ the curve approaches $E_b=1.011\times 2 k_B T_c$ for large interactions which shows that a further increase of the Bose condensation energy is not possible. This gives already a hint that a too strong interaction is not increasing Bose condensation further. In fact we will find in the next chapter within the strong coupling summation that the Bose condensation will decrease with increasing coupling parameter. Here in the weak coupling limit it stays constant. 

A further interesting observation is that for lower densities $\epsilon>0$ there appears a minimal coupling parameter above which only a positive Bose condensation energy is possible. Below this minimal interaction parameter we have a strong imaginary part of the pole (\ref{bcond}) indicating an instable state with a finite lifetime. This minimal interaction parameter as a function of $\epsilon$ is seen in the inset of figure~\ref{i_ey0}. In fact it rises exponentially such that for a given interaction strength we have an upper limit in $\epsilon=-\mu/T$ up to which Bose-Einstein condensation is possible.

\subsection{Inverse selfenergy expansion}

We can describe the strong coupling limit by the proper inversion method as demonstrated in the following. Analogously to the inversion method discussed in chapter II we identify now the order parameter $\phi$ with $x$, the parameter $J$ with $\tilde \epsilon$ and the expansion parameter $g$ with $y_0$. 

The separation line of Bose-Einstein condensation appears for vanishing $\tilde \epsilon$. 
Inverting the small $\tilde\epsilon$ expansion of the function $x(\tilde\epsilon)$ in the second line of (\ref{sys2}) the resulting $\tilde\epsilon(x)$ expansion corresponds to the finite series (\ref{J}). Inserting this into the first line of (\ref{sys2}) gives $y_0^c(x)$ which again is inverted leading to 
\be
x=1+c_1 y_0^c+c_2 (y_0^c)^2+c_3 (y_0^c)^3+...
\label{phaset}
\ee
with
\be
c_1&=&\frac{16 \pi }{3 \zeta\! \left(\frac{3}{2}\right)^{4/3}}=4.657
\nonumber\\
c_2&=&\frac{32 \pi  \left(14 \pi +9 \zeta\! \left(\frac{1}{2}\right) \zeta\!
   \left(\frac{3}{2}\right)\right)}{9 \zeta\! \left(\frac{3}{2}\right)^{8/3}}=8.325
\nonumber\\
c_3&=& \frac{512 \pi  \left(88 \pi ^2+108 \pi  \zeta\! \left(\frac{1}{2}\right) \zeta\! \left(\frac{3}{2}\right)+27 \zeta\! \left(\frac{1}{2}\right)^2 \zeta\!
   \left(\frac{3}{2}\right)^2\right)}{81 \zeta\! \left(\frac{3}{2}\right)^4}
\nonumber\\&=&-14.032
\label{b}
\ee
where we have derived the expansion up to 8th order in $\tilde\epsilon$ in order to achieve sufficient convergence. It has to be remarked that the series possesses diverging terms with alternating signs which nevertheless converge to two limiting curves. 

\begin{figure}
\psfig{file=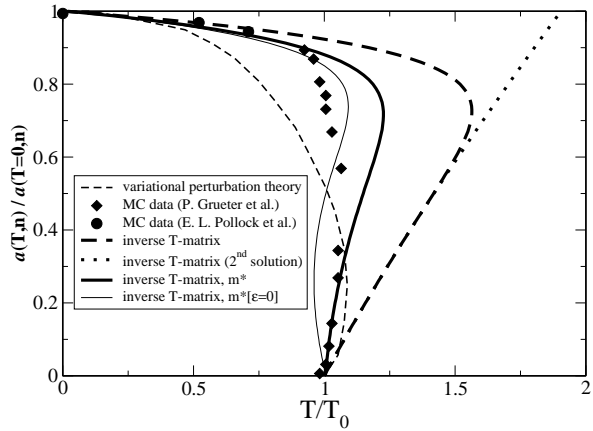,width=9cm}
\caption{The phase diagram of the scaled dimensionless scattering length versus the critical temperature. The MC data are from \protect [\onlinecite{GCL97,PR92}] and the variational improved perturbation theory from \protect [\onlinecite{KSP04,KSP04a}].\label{diag}}
\end{figure}

The results (\ref{phaset}) are plotted in figure \ref{diag}. One of the two limiting curves (dotted line), called $2^{\rm nd}$ solution, corresponds to the naive elimination $x=x[\epsilon(y_0^c)]$ discussed above. The thick dashed line shows the new result due to the inversion method which goes beyond the level of the T-matrix summation of diagrams.  We see that the transition curve bends back and a maximal scattering length occurs below which Bose condensation can happen. It is instructive to compare also the absolute values of the maximal critical parameters where the Bose condensation vanishes. These are presented in table~\ref{tab}. Our maximal critical temperature is too high. In this respect the variational perturbation method  [\onlinecite{KSP04a}] (thin dashed line in figure~\ref{diag}) is nearer to the MC data. This will be improved in the next chapter.

Let us stress again that this inversion procedure is not a simple expansion of $x(y_0^c)$ in powers of $y_0^c$ but primarily an expansion in $\tilde \epsilon(x)$. In fact we had to expand up to second order in $\tilde \epsilon$ in order to obtain the third-order coefficient in (\ref{phaset}), to third order in  $\tilde \epsilon$ in order to obtain fourth order in (\ref{phaset}), etc. This is the reason why we do not get a logarithmic term for the second term as obtained perturbatively in second order in [\onlinecite{AMT01}].

\section{Further improvements}
\subsection{Towards selfconsistency}

Though the figure \ref{diag} describes the data quite well it is mainly 
due to the scaling of the $a$ axis to one. It is more instructive to compare the leading order $c_1$ explicitly. The non-selfconsistent T-matrix 
leads to an increase which is too large by nearly a factor 2 compared with [\onlinecite{BBHLV01,PST01,BBZ00,HK99}]. We can then improve this result by the selfconsistent T-matrix. This is achieved if we replace the free dispersion $p^2/2m$ in the arguments of the Bose function in (\ref{n0}),(\ref{JJ}) and (\ref{aa0}), by the quasi-particle one $\epsilon_p$ which is a solution of the selfconsistent equation
\be
\epsilon_p&=&{p^2\over 2m}+{\rm Re} \Sigma(\omega,p)|_{\omega=\epsilon_p}.
\label{quasi}
\ee
The retarded selfenergy reads [\onlinecite{MLS00}]
\be
&&\Sigma(\omega,k)=\sum_p T_{{k-p\over 2},{k-p\over 2}}(\omega+\epsilon_p,k+p) f_p
\nonumber\\&&
+\sum_{pq} {|T_{{k-p\over 2},{k-p\over 2}-q}(\epsilon_{k-q}+\epsilon_{p+q},k+p)|^2 f_{k-q}f_{p+q}\over \epsilon_{k-q}+\epsilon_{p+q}-\epsilon_p-\omega+i\eta}
\nonumber\\&&
\label{ret}
\ee
with the retarded renormalized T-matrix (\ref{JJ}) in the limit of contact interaction
\be
&&T_{{k-p\over 2},{k-p\over 2}-q}(\omega,k+p)=
\nonumber\\&&-{4 \pi \hbar ^2 a_0/m\over 1\!+\!4 \pi \hbar^2 a_0 \sum_{p'} \left ( {1\over p'^2}\!-\!{1\!+\!2 f_{{p\over 2}\!+\!p'}\over p'^2\!+\!{p^2\over 4}\!-\!{(p\!-\!k)^2\over 2} \!-\!m \omega \!-\!i\eta} \right )}.
\nonumber\\&&
\label{tr}
\ee
Since the medium-dependent scattering length (\ref{aa0}) is determined by the small $k$ expansion and the poles of the different occurring Bose functions appear for small $\tilde \epsilon$ and consequently small momenta, we consider the quasiparticle features only at small momenta. One should note that we are still in the gas-like limit for the quasiparticle dispersion as can be seen from its quadratic behavior. Further improvements require the consideration of the condensate explicitly [\onlinecite{G95}] leading to linear phonon-like dispersions [\onlinecite{SG98}] up to roton minima at higher momenta [\onlinecite{NK06}]. For the present level we restrict our investigation to the quasiparticle renormalization at small momenta and quadratic dispersion. 

The one-particle dispersion relation takes the form 
\be
\epsilon_p&\approx& {p^2\over 2m}+{\rm Re}  \Sigma(0,0)\nonumber\\
&+&\left.\left({p^2\over 2m}{\partial \over \partial {p'^2\over 2 m}}+\epsilon_p{\partial \over \partial \omega'}\right){\rm Re} \Sigma(\omega',p')\right|_{p'=\omega'=0} \!\!+ {\cal O}\left(p^3\right)
\nonumber\\
&\equiv&{p^2\over 2m^*}+\alpha{\rm Re} \Sigma(0,0) + {\cal O}\left(p^3\right)
\ee
with 
\be
\alpha = {1\over 1 - \left.{\partial \over \partial \omega'}{\rm Re} \Sigma(\omega',0)\right|_{\omega'=0}}
\ee
and the effective mass
\be
&&{m\over m^*(\epsilon,y_0)}=\alpha \left ( {1+\left .{\partial\over \partial {p'^2\over 2 m}}{\rm Re}\Sigma(\omega',p')\right |_{\omega'=p'=0}}\right )\nonumber\\
&&=\!\left [ \!1\!-\!\left({\partial\over \partial {p'^2\over 2 m^*}}\!+\!\left .{\partial \over \partial \omega'}\right){\rm Re} \Sigma(\omega',p')\right |_{\omega'=p'=0}\right]^{-1}\!\!.
\label{m}
\ee
The second expression is the selfconsistent one taking into account that the effective mass appears in all expressions of the selfenergy.
The constant selfenergy shift $\alpha {\rm Re}\Sigma(0,0)$ can be included into the definition of the chemical potential in the same way as the mean-field contribution which does not lead to any change in the critical temperature. All what is left from selfconsistency is the effective mass.

\begin{table}
\begin{tabular}{|c|c|c|c|}
\hline
& $(T_c/T_0)_{\rm max}$ & $y_0^{\rm crit}$ &$c_1$
\cr
\hline
&&&
\cr
T-matrix & 1.56 &0.16 & 4.66
\cr
T-matrix (m*($\epsilon$)) &1.22 & 0.16 & 1.32
\cr
T-matrix (m*(0)) &1.07 & 0.16 & -2.13
\cr
Screened T-matrix & 1.56 & 0.32 & 2.33
\cr
MC data \protect [\onlinecite{GCL97}] &1.06 &-& 0.34$\pm$0.08  
\cr
MC data \protect [\onlinecite{HK99}] &-&-&2.30$\pm $0.25
\cr
MC data \protect [\onlinecite{AM01}] &-&-&1.32$\pm $0.02
\cr
MC data \protect [\onlinecite{KPS01}] &-&-&1.29$\pm $0.05
\cr
\hline
\end{tabular}
\caption{The maximal critical temperature, the critical coupling parameter where the Bose condensation vanishes and the linear coefficient from the critical temperature according to figure~\protect\ref{diag} for the three different approximations used in the text.\label{tab}}
\end{table}

It is not difficult to see from (\ref{n0}) and (\ref{x0}) that in the second line of (\ref{sys2}) an additional factor 
$m/m^*$ has to appear
\be
{y_0^{c}}&=&-{1\over 4 g'(\tilde \epsilon)}
\nonumber\\
x&=&{g(0)\over g(\tilde \epsilon)} {m\over m^*(\tilde \epsilon, y_0^c)}.
\label{sys3}
\ee

For our present discussion we have used only the first part of the selfenergy (\ref{ret}). The system of equations (\ref{sys3}) is solved and the resulting curve approaches the MC data as can be seen in figure~\ref{diag} (thick solid line). Especially the maximal critical temperature 
is nearer to the MC data as can be seen in table~\ref{tab}. Now the effective mass depends on $\tilde \epsilon$. This leads to a delicate numerical balance as illustrated by the curve for the effective mass at fixed $\tilde\epsilon=0$ which gives the best overall agreement with the data points but shows a negative $c_1$. Calculations with higher-order approximation of the selfenergy are therefore certainly necessary to achieve a sufficient convergence.


\subsection{Screened ladder diagrams}

We can now further improve the many-body approximation by using a ring summation of diagrams for the potential of the T-matrix. This corresponds to the screened ladder approximation such that the bare potential in the T-matrix is replaced by a screened one  [\onlinecite{zkkkr78}]. In order to maintain separability of the T-matrix we consider this screened potential in separable form as a result of the screening
\be
\tilde V_{pp'}=V_{pp'}-\sum\limits_{\bar p} \tilde V_{p{\bar p}} {f_{\frac p 2 +{\bar p}}-f_{\frac p 2 -{\bar p}} \over \varepsilon_{\frac p 2 +{\bar p}}-\varepsilon_{\frac p 2 -{\bar p}}-\omega +i \eta} V_{{\bar p} p'}
\ee
which is solved yielding
\be
\tilde V_{pp'}&=&{\lambda g_pg_{p'}\over 1+\lambda \Pi(\omega,p)}
\nonumber\\
\Pi(\omega,p)&=&\sum\limits_{\bar p} g_{{\bar p}}^2 {f_{\frac p 2 +{\bar p}}-f_{\frac p 2 -{\bar p}} \over \varepsilon_{\frac p 2 +{\bar p}}-\varepsilon_{\frac p 2 -{\bar p}}-\omega +i \eta}. 
\ee
We see that the screening corresponds to a scaling of the coupling constant in the T-matrix (\ref{t-mat}) 
\be
\lambda \to {\lambda\over 1+\lambda \Pi(\omega,p)}
\ee 
and the same solution as (\ref{JJ}) appears but with a new momentum- and energy-dependent coupling. We can now follow the same procedure as outlined above in using a $\beta$-dependent coupling constant to perform the contact potential limit $\beta \to \infty$. For this aim we need the small momentum expansion
\be
&&\!\!\!\!\lim\limits_{\beta \to \infty}\!\Pi\left({p^2\over 2 m},p\right)=-{m\over 2\pi^2\hbar^3 \beta^4}\!\int\limits_0^\infty\!\! dp f_p\!+\!i {m p f_0  \over 8 \pi \hbar^3 \beta^4} \!+\!{\cal O}(p^2).
\nonumber\\
&&
\ee
The resulting formula analogous to (\ref{aa0}) reads now
\be
{a\over a_0}={1+3 f_0\over 1-{2 a_0 \over \pi \hbar}\int\limits_0^\infty d{\bar p} f_{\bar p}}
\label{aa1}
\ee
which leads to
\be
{y_0^{c{\rm s}}}&=& -{1\over 2 g'(\tilde \epsilon)}
\nonumber\\
x&=& {g(0)\over g(\tilde \epsilon)}
\label{sys4}
\ee
instead of (\ref{sys2}). Therefore an additional factor $1/2$ appears in the linear coefficient (\ref{b}),
\be
c_1^{\rm s}=\frac{8 \pi }{3 \zeta\! \left(\frac{3}{2}\right)^{4/3}}=2.33,
\ee
a result in remarkably good agreement with [\onlinecite{BBHLV01,PST01,BBZ00}] and the MC data [\onlinecite{HK99}]. The critical values are found in table~\ref{tab}. The screened ladder approximation leads to twice the maximal critical coupling strength but the same maximal critical temperature compared to the T-matrix approximation. We note that further improvements should be possible again by considering the selfconsistency, i.e. the effective mass.

\section{Summary}

We have discussed interacting Bose systems with a repulsive interaction. The appearance of the Bose-Einstein condensation becomes dependent on the interaction parameter. We derive the same condition for Bose-Einstein condensation both from the pole of the T-matrix and from the observation that the medium-dependent scattering length is diverging at the Bose condensation. This condition is used to produce a perturbation series of higher order than the original T-matrix summation adopting the well known method of inversion. With the help of this method we obtain a strong coupling result out of the weak coupling T-matrix summation. We are able to describe the nonlinear behavior of the critical temperature. The critical temperature rises linearly with the interaction parameter in leading order reaching a maximal value and is reduced for higher interaction strength. At an upper specific interaction parameter Bose condensation is not possible anymore. The system is too repulsively correlated for allowing condensation. 

We improve the original weak-coupling series further by using screened vertexes which reproduce the leading order coefficient of the MC data. Further improvements towards the nonlinear curve of the MC data and the variational perturbation theory are possible by the selfconsistency. Here we present as a first step an effective mass calculation inside the T-matrix and show that the inversion method leads then to a better description of the data.


\bibliography{bose,kmsr,kmsr1,kmsr2,kmsr3,kmsr4,kmsr5,kmsr6,kmsr7,delay2,spin,refer,delay3,gdr,chaos,sem3,sem1,sem2,short}
 
\end{document}